\input psfig.sty 

 \documentstyle{l-aa} 
\begin{document} 
\thesaurus{05.01.1 08.10.7} 
 
\title{Hipparcos Positioning of Geminga : How and Why. $^{\spadesuit, \clubsuit}$}

\author{P. A. Caraveo\inst{1}, M.G. Lattanzi\inst{2}, G. Massone \inst{2}, R. P. 
Mignani\inst{3}, V.V. Makarov\inst{4}, M.A.C. Perryman\inst{5} and G. F. 
Bignami\inst{6,1}}

\offprints{P. Caraveo, pat@ifctr.mi.cnr.it} 

\institute{$^{1}$ Istituto di Fisica Cosmica del CNR, Milano, Italy \\ 
           $^{2}$ Osservatorio Astronomico di Torino,  10025 Pino Torinese, Italy \\
           $^{3}$ Max-Plack-Institute f\"ur Extraterrestrische Physik, Garching, 
Germany \\         
           $^{4}$ Copenhagen University Observatory, 1350 Copenhagen K, Denmark \\
           $^{5}$ Astrophysics Division, ESTEC, 2200AG Noordwijk, The Netherlands \\
           $^{6}$ Agenzia Spaziale Italiana, Via di Villa Patrizi 13, Roma, Italy
}   

\date{Received ....; Accepted .... } 
\maketitle 
\markboth{P.A. Caraveo et al.: Hipparcos Positioning of Geminga}
\noindent

$\spadesuit$ Based on Observation with the ESA Hipparcos satellite.\\ 
$\clubsuit$ Based on observations with the NASA/ESA Hubble Space Telescope.   


\begin{abstract} 
Accuracy in the absolute position in the sky is one of the limiting factors 
for pulsar timing, and timing parameters have a direct impact on the 
understanding of the physics of Isolated Neutron Stars (INS). We report 
here on a high-accuracy measurement of the optical position of Geminga 
($m_{v}=25.5$), the only known radio-quiet INS.  The procedure combines 
the Hipparcos and Tycho catalogues, ground-based astrometric data,and 
Hubble Space Telescope (HST) Wide Field Planetary Camera (WFPC2) 
images, to yield Geminga's absolute position to within $\sim 40~mas $ 
(per coordinate). Such a 
positional accuracy, unprecedented for the {\it optical} position of a 
pulsar or 
an object this faint, is needed to combine in phase $\gamma$-ray photons 
collected over more than 20 years, i.e. over 2.5 billions of star' 
revolutions. Although quite a difficult task, this is the only way to 
improve our knowledge of the timing parameters of this radio silent INS.

 \vspace {5pt} 


  Key~words: space astrometry; pulsar; Geminga{}.

\end{abstract}

\section{Introduction.} 
Geminga is the only known example of a bona fide pulsar that cannot 
rely on well established radio timing techniques. The radio 
silence of the source implies the impossibility to measure its position 
and distance as inferred from the optimization of the radio timing 
parameters and from the dispersion measure of the radio pulses. 
To measure position, distance and timing parameters of Geminga, all 
other branches of astronomy had to be exploited in a 20 y long chase, 
recently summarized in Bignami and Caraveo (1996 and references therein).  
Briefly stated, the source was discovered in high energy $\gamma$-rays by the 
SAS-2 satellite in 1972,  and studied in more detail by the COS-B 
mission. An X-ray counterpart has been proposed in 1983 and an optical 
one, refining the position, in 1987/88. However the breaktrough came 
with the discovery of the 237 msec periodicity in the ROSAT data. 
Finding the same periodicity in the contemporary data of EGRET, as well 
as in old archival COS-B and SAS-2 data, yielded the value of the period 
derivative and thus of the object's energetics. The discovery of the 
proper motion of the optical counterpart confirmed the optical 
identification and provided the absolute position of Geminga to within 
1". Next came the measure of the source parallactic displacement, 
yielding a precise measure of its distance.
Our knowledge of the Geminga pulsar is now good enough to warrant, 
"honoris causa", inclusion in the radio pulsar catalogue Taylor, 
Manchester and Lyne (1993). In it, the 
radio silent Geminga stands out for the remarkable accuracy achieved in 
the measure of its parameters.  For example, the period derivative of 
Geminga is known with an accuracy greater than that of the Crab. Such 
an accuracy is mainly due to the very stable behaviour of this $3 10^{5} y $
old neutron star, which does not seem to be affected by glitches. Indeed, 
Mattox, Halpern and Caraveo (1996) claim that, during the first 3 years of 
coverage with EGRET, every pulsar revolution can be accounted for. This 
means that the $\gamma$-ray photons, collected in week-long observing periods 
taken several month apart over a span of years, can be aligned in phase 
to form the well-defined, spiky light curve seen in high energy $\gamma$-rays.   
Had Geminga been a glitching pulsar, like Vela or Crab, it would have 
been impossible to obtain a satisfactory (and accurate)  long-term 
solution. \\Moreover, the growing high energy coverage offers the 
possibility to refine the knowledge of this source's timing parameters,  
including the second period derivative and, thus, the neutron star 
braking index.  However, to take full advantage of the potential offered 
by $\gamma$-ray astronomy, very accurate values of the absolute source 
coordinates are required to determine the barycentric arrival time of 
each photon.  Any error in the source coordinates affects the accuracy of 
the corrected arrival times and thus hampers the global accuracy of the 
procedure. Mattox, Halpern and Caraveo (1996) have shown that the 
uncertainty in the source absolute positioning can induce an error in the 
barycentric correction up to 2.3 $\delta_{e}$ msec, where $\delta_{e}$ is 
the source 
positional uncertainty (in arcsec) projected on the plane of the ecliptic. 
Thus, with a sensitivity to phase errors of $<10^{-2}$ over a period of 237 
msec, the presently available positional accuracy of $\sim 1"$ is the limiting 
factor for Geminga's photon timing analysis. As mentioned before, 
high-energy $\gamma$-ray photons have been collected unevenly over the years:  
during 1972-1973 by the SAS-2 satellite, during 1975-1982 by the COS-B 
mission and again from 1991 by the Compton Gamma-Ray Observatory. 
To lock in phase $> 20$ years of $\gamma$-ray data an improvement in positional 
accuracy of at least one order of magnitude is required. To achieve this challenging goal
we 
have exploited the angular resolution of HST in conjunction with the 
accuracy of the Hipparcos reference frame (ESA, 1997).

\section{From Hipparcos to HST}
However, to tie Geminga into the Hipparcos system, one has to overcome 
the 16-18 magnitude gap between the bright stars used as a primary 
reference and our $m_{v}=25.5$ target, an impossible dynamical range for a 
single instrument.  Moreover, one has to link astrometric images 
covering a field of view of $\sim$1 square degree to those obtained by the 
WFPC2 of HST, 14,000 times smaller in area. \\The principle of our method, described in 
detail by Caraveo et al. (1997), 
is to cover the field of interest with images 
of increasingly smaller field of view and deeper limiting magnitude. The 
transfer of the reference system from one step to the next  is based on 
sets of stars of intermediate magnitude. 
A similar procedure has been successfully applied (e.g., Chiumiento et al, 1991; 
and Zacharias et al, 1995) for 
the determination of absolute positions of optical counterparts of 
extragalactic radio-sources.  However,  these cases were less critical, as 
the differences in magnitude were only 8-10 and two or three 
intermediate steps were usually sufficient to bridge the gap between the 
reference stars and the targets.   We note that the FK4 coordinates of 
the 16 mag Crab pulsar were obtained by MacNamara (1971) with a two step 
procedure. \\In the case of the 9 magnitude fainter Geminga, we had to use a new 
five-step procedure starting with astrometric plates taken at the 
Osservatorio Astronomico Torino (OATo), and  ending with images taken 
with the Planetary Camera 2 on HST (see Figure 1 for a comparison of the 
two images). 
The observational material actually used in this work is listed in Table1.
To calibrate the astrometric plates we have used the best optical 
reference frame in the sky which is now provided by the Hipparcos and 
Tycho Catalogues. These catalogues have been constructed in such a 
way that the Hipparcos reference frame coincides, to within limits set by 
observational uncertainties, with the International Celestial Reference 
System (ICRS). The latter system is defined by the adopted positions of 
several hundred extragalactic radio sources. It supersedes, although it is  
consistent with, the optical reference frame defined by the FK5 
catalogue, which was formally based on the mean equator and 
dynamical equinox of J2000. While the Hipparcos Catalogue lists 118,000 
stars as faint as $m_{v}=12.4$ whose positions and annual proper motions are 
known to typically 1~mas at epoch J1991.25, the Tycho Catalogue 
contains more than one million objects, the astrometric parameters of 
which have been measured with a median precision of 25~mas. 
With about 3 stars per square degree the Hipparcos Catalogue, although 
of vastly superior precision, is not suitable for our purpose, which can be 
better fulfilled by the relatively less accurate but denser Tycho 
Catalogue. However, since the proper motions listed in the Tycho 
Catalogue are accurate only to approximately 25~mas/yr, we have 
propagated the Tycho positions (epoch J1991.25) to the OATo 38~cm 
refractor plate epoch (J1984.19) using the proper motions provided by 
the PPM catalogue (R\"oser and Bastian, 1993). This is an all-sky 
list of 378,910 stars referred to the FK5 system and provides proper 
motions with a typical accuracy of 4~mas/yr. Consequently, we used a 
subset of the Tycho objects also listed in the PPM catalogue as our 
primary reference frame. 
On our first astrometric plate (left panel in Fig.1), 19 Tycho stars in common 
with the PPM define the primary reference, while a grid of 17 fainter 
stars is used as the secondary reference frame. This grid of fainter stars 
acts as the  "primary reference frame" on the plates taken with the 105 
cm astrometric reflector, defined as step 2 in Table 1.  The procedure is 
then repeated to transfer the reference frame from step 2 to step 3, 
from 3 to 4, and from 4 to 5, ending with a grid of reference stars 
suitable for the calibration of the WFPC2 frame (right panel of Fig.1).
The optical and mechanical design of the OATo astrometric telescopes
minimizes optical distorsions over fields of view even larger
than the present ones. Indeed, no plate
modeling residual effects (see i.e. Chiumiento et al, 1991; Lattanzi et al, 1991) can be seen 
in our data. \\  The geometric distorsions in HST/PC data have been corrected following the 
standard procedure described in  Holtzmann et al (1995).
The accuracy of our procedure can be evaluated from the values of $\epsilon_r$, 
the star centering errors which are typical of each telescope/detector combination, 
and $\epsilon_{tr}$, the errors due to the least square procedure used to generate 
the secondary stars grid. These are also given Table 1.

\begin{table*}
\caption{{\it The stars in the secondary grid of step n (5th column) become 
the primary grid of step n + 1 (4th column). 
Since some of the secondary grid stars measured at
step n could not be satisfactorily measured on the image of
step n+1 (owing to, e.g., saturation, duplicity, etc.) the numbers
in the two columns can be different.
For each step we also give the centering ($\epsilon_r$) errors as well as the 
uncertainties arising from the frame transfer procedure
($\epsilon_{tr}$). The values for step 5 are educated guesses and not direct estimates, as 
for the other data. The limiting magnitude for each frame is listed in the last column. 
The value given for step \#4 refers to the stack of 10 frames of 15 minutes, each of them providing a
limiting magnitude of 24.5 
}} 

\begin{center} 
\begin{tabular} {|c|l|c|c|c|l|l|c|} \hline
  
{step} & {telescope/detector} & { field of view} &  {primary grid} &  {secondary grid} &  {$\epsilon_r$} &  {$\epsilon_{tr}$} &
{\em mag} \\
{[N. images]} & & {[epoch]} & {} & {} &  {} & {} &  \\ \hline

step 1 & OATo 38cm refractor  & 70' x 70' & 19 Tycho/PPM & 17  & 0".102   & 0".044  & 14  \\
2 & plate,  30"/mm & [1984.19] & & & & & \\
step 2 & OATo 105cm reflector & 30' x 30' & 16 & 28  & 0 .061   & 0 .020 & 17 \\
2 & plate,  20".7/mm & [1984.19] & & & &  & \\                             
step 3 & OATo 105cm reflector &  9' x 10' & 26 & 21  & 0 .026   & 0 .011 & 19.5  \\
2 & CCD,  0".48/px & [1996.13] & &  &  & & \\
step 4 & ESO NTT 3.5m, SUSI & 2.5' x 2.5'& 16 & 10 & 0 .015   & 0 .008  & 26 \\
1 & CCD  0".13/px& [1992.86] & & &  & & \\
step 5 & HST 2.4m,  PC2 & 35" x 35" & 10 & Geminga & [0 .005] & [0 .003] & 26 \\
1 & CCD, 0".046/px & [1995.21] & & & & & \\ \hline
\end{tabular}
\end{center}
\end{table*}

\begin{figure*}
\centerline{\hbox{
\psfig{figure=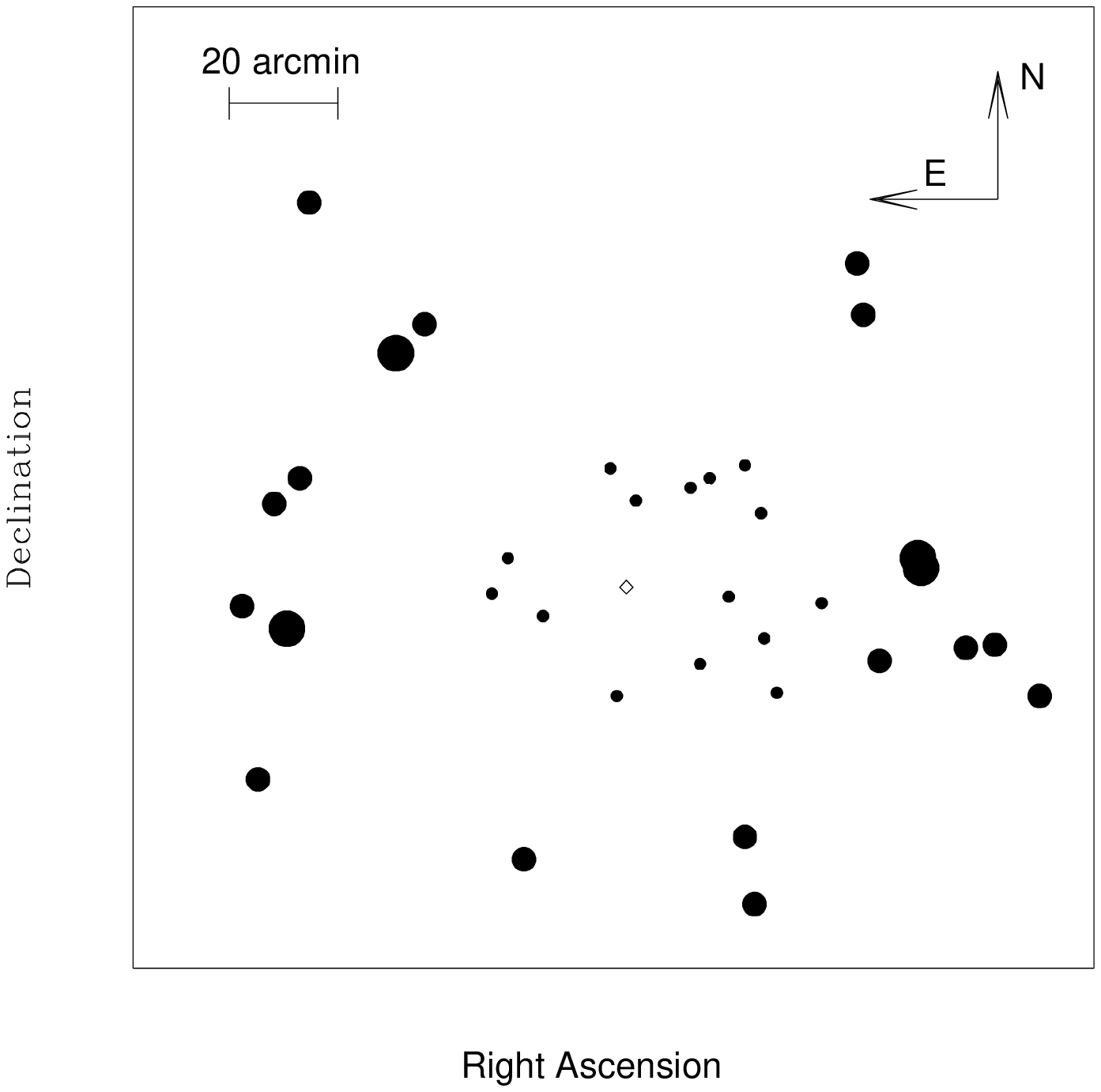,height=8cm,clip=}
\psfig{figure=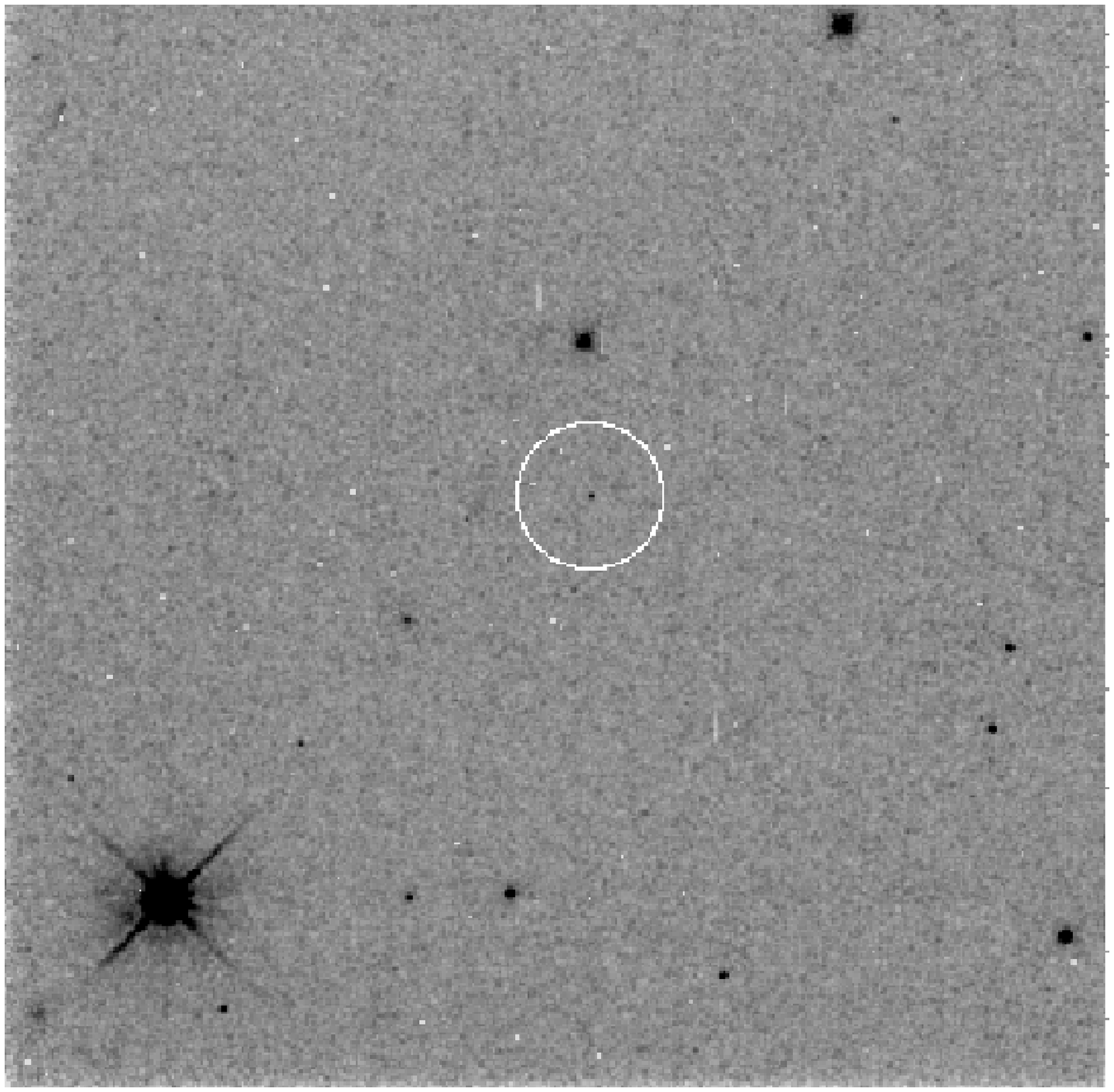,height=8cm,angle=137,clip=}
}}
\caption{{\em Comparison between the first and the last step of our procedure.
Left: schematic representation of the 70' x 70' astrometric plate taken at the 
OATo. Hipparcos stars are marked with large filled circles, Tycho stars with
medium circles, while the small circles in the central region of the plate 
identify the stars used as secondary reference frame. The diamond in the center 
gives the position and actual dimension of the HST image, which represents the 
last step of our chain. Right: 35" x 35" Planetary Camera image of the field 
of Geminga obtained 
with a 4400 sec exposure through filter 555, roughly equivalent to V. 
Geminga is shown inside a white circle.}}
\end{figure*}

The resulting 
position of Geminga  at epoch=1995.21 is\\
$\alpha = 6^{h} 33^{m} 54.1530^{s}$,  $\delta= 17^{o}  46' 12.909"$;  
these coordinates are in the Hipparcos 
reference frame and, therefore, in the ICRS.\\To this value we attach the error  $\sqrt{\epsilon_{HST}^2 + \epsilon_{sys}^2 + 
\epsilon_{Tycho}^2}$ , where\\ 
$\bullet \epsilon_{HST}$ is the centering error $\epsilon_{r}$ of the HST image, 
which is conservatively taken 
as 1/10 pixel (~0."005);\\
$\bullet \epsilon_{sys}$ results from the combination of the errors induced by the 
least square adjustments used, at each step, to generate the lists of 
secondary stellar grids $\epsilon_{sys}$ (Eichhorn and Williams, 1963). Following  
(Lattanzi et al, 1997), for m stars with 
average centering error $\epsilon_{r}$, such transformation errors can be written 
as $\epsilon_{tr} \sim \sqrt{3}
\times \epsilon_{r}/\sqrt{m}$, where 3 is the number of free parameters in the 
linear plate-to-field transformation procedure. Table 1 gives 
the centering errors $\epsilon_{r}$ and the frame transfer ones $\epsilon_{tr}$ 
for each step of 
our procedure. Adding in quadrature the $\epsilon_{tr}$'s one obtains 
$\epsilon_{sys}$ 
0."050. Since this procedure has been independently performed for 
the two images available for steps 1 through 3, a reduction of $\sqrt{2}$ is 
to be applied to the single solution  $\epsilon_{sys}$.\\
$\bullet \epsilon_{Tycho}$ measures the precision with which we can register 
the ensamble 
of 17 stars, defining the secondary reference frame on the refractor 
image,  relative to that realized by the 19 Tycho/PPM stars used as a 
primary reference on the refractor image. In our case 
$\epsilon_{Tycho} \sim \sqrt{3} \times \sigma_{Tycho}/\sqrt{19}$, where
$\sigma_{Tycho} \simeq 0."032$ is the mean error (per coordinate) at the 
epoch of the refractor plate accounting for both 
the average errors in the Tycho positions (~15 mas for our sample) 
and the uncertainty on PPM proper motions propagated over 7 
years. 
Therefore, the final error to be attached to the ICRS position of Geminga 
is ~40 mas per coordinate.\\ Thus, the combined use of HST and Hipparcos yielded a 25 fold 
improvement  in the source absolute positioning.

\section{On the Use of an Accurate Position}

Among the pulsars seen in the optical or as high energy emitters, 
Geminga's positional accuracy becomes the best so far, better than that 
of the 9 magnitudes brighter Crab pulsar (McNamara, 1971).
This improved position, in conjunction with the HST measure of the 
proper motion and parallax (Caraveo et al., 1996), will allow the accurate 
calculation of the 
barycentric arrival time of all the available $\gamma$-ray data and the locking in 
phase over more than 20 yrs of data from three separate space missions. 
It will thus be possible to measure the period second derivative and, 
hence, the braking index $n= \ddot{\nu} \nu / \dot{\nu}^{2}$ of this neutron star. 
This quantity, expected to be 3 for magnetic dipole braking, has been 
only measured so far for young objects such as Crab (Lyne et al. 1988; $n= 2.5 
\pm .01$), PSR 0540-69 (Guiffes et al.199; $2.8 \pm .01$), PSR 1509-58 (Kaspi et al.,1994; 
$n=2.0 \pm .2$), and for the slightly older PSR 0833-45 (Lyne et al. 1996; 
$n=1.6 \pm .3$).   \\While the results obtained for the three very young objects are not too 
far from the expectations, the braking index recently measured for the 
$\sim 10^{4} yr$ Vela pulsar is definitely lower. Hence, the importance to exploit 
the stability of the $3~10^{5}~ y$ old  Geminga to considerably enlarge the 
pulsar age sampled for braking index determination. The task is a challenging one : 
for a canonical braking index of 
3, a $\ddot{\nu}$  of $2.7 10^{-26} s^{-3}$  is expected. This value is three order
of magnitude 
smaller than that recently measured for Vela and 6 order of magnitude 
smaller than that of the Crab. However, such tiny value of $\ddot{\nu}$  
is within 
reach of the SAS-2, COS-B and CGRO data.   \\What makes Geminga suitable for the measurement of the frequency 
second derivative?  If we order known pulsars according to increasing 
values of $\ddot{\nu}$ (expected for a braking index of 3), Geminga does not 
come out prominently. Not less than 40 pulsars  have hypothetical 
$\ddot{\nu}$ bigger 
than the value expected for Geminga. However, for all of them, these 
values are not measurable, because the errors on $\nu$ and $\dot{\nu}$ 
are too big. 
What singles out Geminga is the possibility to reduce such errors by 
phasing together 20 years worth of $\gamma$-ray data, now that the source 
positional accuracy is no longer a limiting factor. Thus, once again, 
Geminga appears to be in a special position amongst Isolated Neutron 
Stars. What matters most here is the source instrinsic stability, which 
made it worthwhile to devote a dedicated effort to the accurate 
measurement of its absolute position.

\section {Acknowledgments}
We are grateful to the Hipparcos and Tycho collaborations who made
the positions available to us prior to the official release of the catalogues.

\end{document}